\documentclass{jimis-final-en}
\usepackage{array}

\usepackage{graphicx}
\usepackage[T1]{fontenc}
\usepackage[utf8]{inputenc}
\usepackage[francais]{babel}
\usepackage{caption}
\usepackage[left,modulo,pagewise]{lineno}
\usepackage{amssymb,amsmath,amsfonts}
\usepackage{float}
\usepackage{multirow}
\usepackage[]{algorithm2e}
\usepackage{url}
\usepackage{tikz}
\usepackage{pgfplots}

\usepackage{hyperref}
\usepackage{color}

\title{Active learning in annotating micro-blogs dealing with e-reputation}
\author[1,3,4]{Jean-Val\`ere Cossu}
\author[2,3]{Alejandro Molina-Villegas}
\author[2]{Mariana Tello-Signoret}
\affil[1]{Vodkaster, France} 
\affil[2]{Conacyt, Mexico} 
\affil[3]{Université d'Avignon, France} 
\affil[4]{My Local Influence, France} 

\corrauthor{jvcossu@gmail.com}

\doi{10.18713/JIMIS-010917-3-2}{}
\review{05/07/2017}{05/09/2017}
\publication{3}{2017}

\issue{Digital Contextualization}
\editors{Frédéric Lebaron, Brigitte Le Roux, Fionn Murtagh, Evelyn Ruppert}

\begin{document}

\maketitle


\abstract{Elections unleash strong political views on Twitter, but what do people really think about politics? Opinion and trend mining on micro blogs dealing with politics has recently attracted researchers in several fields including Information Retrieval and Machine Learning (ML). Since the performance of ML and Natural Language Processing (NLP) approaches are limited by the amount and quality of data available, one promising alternative for some tasks is the automatic propagation of expert annotations. This paper intends to develop a so-called active learning process for automatically annotating French language tweets that deal with the image (i.e., representation, web reputation) of politicians. Our main focus is on the methodology followed to build an original annotated dataset expressing opinion from two French politicians over time. We therefore review state of the art NLP-based ML algorithms to automatically annotate tweets using a manual initiation step as bootstrap. This paper focuses on key issues about active learning while building a large annotated data set from noise. This will be introduced by human annotators, abundance of data and the label distribution across data and entities. In turn, we show that Twitter characteristics such as the author's name or hashtags can be considered as the bearing point to not only improve automatic systems for Opinion Mining (OM) and Topic Classification but also to reduce noise in human annotations. However, a later thorough analysis shows that reducing noise might induce the loss of crucial information.}

\keywords{Opinion Mining, Online Reputation Monitoring, Active Learning, Machine Learning, Human Annotation, Methodology, Sentiment Analysis, Topic Categorization, Natural Language Processing}

\section{Introduction}
In the last decade, there has been a historical change in the way we express our opinion. In a world of online networked information, people are getting used to talk about anything and everything on a multitude of participative social media: forums, reviews, blogs, micro-blogs, etc., user-generated contents in the form of reviews, ratings and any other form of opinion, should be dealt with OM,~\cite{PAK10}. Usually, it is a positive or negative judgment towards a product, formulated by an explicit vote score between one and five stars and/or implicitly by means of natural language (e.g., ``I like the speed of this printer.''),~\cite{Hu:2004}. Recently, using human labeled datasets, the SemEval challenges included tasks about Aspect Based Sentiment Analysis,~\cite{Semeval15} using words, terms and sentences as they are naturally expressed in reviews and tweets.

Since information control has moved to users, OM on micro-blogs such as Twitter has also become very popular to predict future trends. Afterwards, each act of a public entity is scrutinized by a powerful global audience,~\cite{jansen2009twitter}. Therefore, OM had then been used in broader and more difficult contexts such as reputation and politics,~\cite{Wang:2012}. This led to the creation of an emerging research trend towards Online Reputation Monitoring,~\cite{Burton2011}. However analyzing reputation about companies and individuals is a challenging task requiring a complex modeling of these entities (e.g. company, politician). Moreover in the case of tweets there are no explicit ratings to be directly used in an opinion processing. This explains the need for new Reputation Monitoring tools and strategies which also become an interesting way to process large amounts of opinions about various kind of entities,~\cite{Malaga2001}.

Currently, market research employing user surveys is typically performed and traditional Reputation Analysis, ~\cite{glance2005deriving,hoffman2008online} is a costly task when done manually. Processing large amounts of reputation data is a real challenge not only to deal with specific requirements in Information Retrieval or OM, but also to understand important issues in political science,~\cite{gerlitz2013mining,boyadjian2014twitter}. Politics have already been addressed in previous works but mostly in English, German or Spanish,~\cite{kato2008taking,O'Connor:2010,metaxas2011not,park2011politics,Wang:2012,jungherr2012pirate,villena2013tass,Hendricks2014,pla2014political} and more recently with Bulgarian,~\cite{smailovic2015monitoring}. As far as we know, nothing in French has been done from a machine learning perspective until now. 

The work presented in this paper is oriented towards the extraction of opinions together with their target aspects on French political tweets focusing on the two main candidates in the last presidential election in France, in May 2012. This work involved academics as well as industrial partners, including end users (politics researchers) who have been involved in the whole process (from design to evaluation). In contrast to previous research, the scientific contribution is threefold.

\begin{itemize}
  \item Firstly, we collaborate with experts in political science in order to design a full annotation framework and usage scenarios. This will lead to an annotated seed dataset with the involvement of specialists in political science. The annotations are aspect-oriented polarity for reputation. In other words, the opinion expressed on a specific aspect is linked to a dedicated attribute of the entity.
  \item Secondly, we develop dedicated automated classification techniques able to deal with short texts and aspect-oriented opinion statements related to French politics. Our approach relies on automatic propagation of the reduced set of expert annotations we just described among larger collections of tweets.
  \item Thirdly, we intend to study the impact of automatic label proposal on the annotator assessments and investigate the classification performances. Our propagation approach deals with three key issues about active learning while building a large annotated data set:
    \begin{itemize}
    	\item Identify and remove noise introduced by human annotators, 
    	\item Use data abundance,
    	\item Harmonize label distribution across data and entities,~\cite{xu2007incorporating}.
    \end{itemize}
\end{itemize}

The rest of the paper is organized as follows. Section~\ref{sec:state} provides an overview of related works. In Section~\ref{sec:platform} we detail the annotation platform and give basic statistics of the first annotated set. We then study the main characteristics of crowd-sourced annotations about politics in Section~\ref{sec:data}. In Section~\ref{sec:classif} we propose a new pseudo-active learning algorithm for bias correction to improve the quality of annotations and the automatic annotation procedure to increase the final amount of labeled data. Section~\ref{sec:results} introduces use cases evaluation of our algorithm. Finally, we conclude and give some research directions.

\section{Related work}
\label{sec:state}

\subsection{Tweets mining}
Previous works on reputation monitoring in tweet collections and streams have been made to extract sets of messages requiring a particular attention from a reputation manager,~\cite{Amigo2013}. For example, recent contributions to this issue on Twitter data have been done in the context of the 2013-14 editions of Replab \footnote{\url{http://www.limosine-project.eu/events/replab2013}} and TASS \footnote{\url{http://www.sepln.org/workshops/tass/2014/tass2014.php}} challenges where the lab organizers provide a framework to evaluate Online Reputation Management systems on Twitter. 

Reputation polarity is substantially different from standard sentiment analysis, since both author, facts and opinions have to be considered. The goal is to find what implications a piece of information has on the reputation of a given entity regardless of whether the message contains an opinion or not (i.e. news just factually reporting wrong governance decision). To illustrate, if ten humans disagree on the sentiment of a given text, it  then issues if what is acceptable or relevant for one individual is the same for others. Multilingual aspects, cultural factors and context awareness are among the main challenges of sentiment natural language text classification when dealing with reputational micro-blogs.

Furthermore, topic detection is used to guess the topic of the text or the aspect linked to the opinion with two possibilities: one among those of a predefined set of categories or classes, so as to be able to assign the reputation level of the company into different facets, axes or points of view of analysis. Another employing users networks and text similarities to build message groups and consider the topic as the concept expressed by the key features (terms extracted) of each group. Nevertheless, in micro blogging, due to the 140 characters limit, messages are often allusive with few words making both tasks harder.

\subsection{Data building}
Crowd-sourcing is an increasingly popular and collaborative approach for acquiring research annotated corpora with the idea of collecting annotations from volunteer contributors, this is an advantage over expert-based annotation. Although designing such a dataset of training examples has proven quite an interesting challenge,~\cite{Amigo2013},~\cite{villena2013tass}, it is still expensive and relatively inaccurate. The background literature,~\cite{walter2013text} focuses on central points which describe a current research issue. Indeed, although the use of paid-for crowd-sourcing approach is intensifying\footnote{With the emergence of platforms such as Amazon Mechanical Turk \url{www.mturk.com}}, the reuse of annotation guidelines, task designs, and user interfaces between projects is still problematic, since these are usually not available for the community despite their important role in result quality. Moreover, the cost to define a single annotation task remains quite a substantial
challenge for crowd-sourcing projects. 

Literature is also full of innovative approaches about definition of crowd-sourcing success, especially on how to evaluate the results and the application of text mining approaches. Much recent researches focused on the reliability and applicability of crowd-sourcing annotations for NLP,~\cite{wang2013perspectives}. Previous works using so-called active learning,~\cite{settles2012active} have been done to automatically build high-quality annotated datasets on twitter monitoring,~\cite{carrillo2014orma}. Most part of research projects leave behind them a small annotated corpus and a large amount of unlabeled data. The small data set can be used as bootstrapping for systems,~\cite{di2004bootstrapping} but how can we make use of the remaining unlabeled set ? The idea is to utilize the unlabeled examples by adding labeled data which has been well studied in the last decade,~\cite{blum1998combining,mccallumzy1998employing}. 

In our case, as manual annotation is a costly work, we use state-of-the-art approaches to build and improve a dataset. Text mining is then not only applied to handle the issue of semi-supervised annotation but also to fulfill an optimal semi-supervised selection of the messages we want to submit for manual annotation. To answer these key issues, we have designed a protocol which aims to automatically annotate tweets and extract semantic relationships between the expressed polarity and the aspect. In addition to a dataset, we also provide a full open-source annotation platform \footnote{\url{http://dev.termwatch.es/~imagiweb/index.php}} and its design. This design comprises different processes such as data selection, formal definition and instantiation of the reputation. 

\section{Crowd-sourced annotation stage}
\label{sec:platform}
\subsection{Annotation platform for E-Reputation Analysis of tweets in French}
\begin{figure}[!htb]
  \centering
  \includegraphics[ width=0.430 \textwidth ]{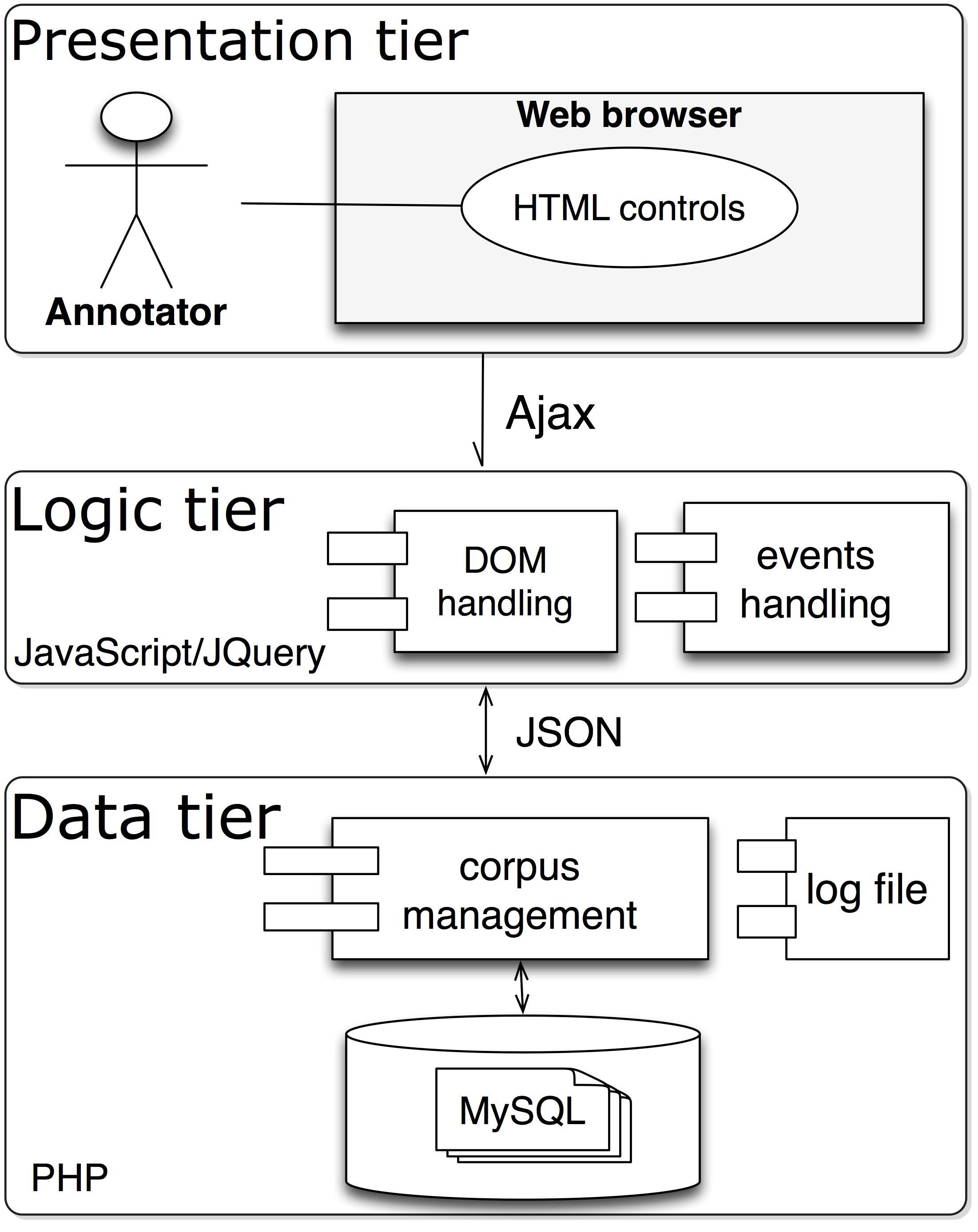}
  \caption[System architecture.]{\label{fig_arch} System architecture.}
\end{figure}

To analyze the public image of French politicians in Twitter we designed an annotation platform where users are given tweets and are asked to first identify the opinion passage; then to assign it to a polarity and finally to identify its specific aspect target. Our Web architecture, shown in Figure \ref{fig_arch}, is based on the three-tier models which allow a quick adaptation to any annotation needed because mostly the top-most level source code must be modified.

\begin{figure}[!ht]
  \centering
  \includegraphics[width=1.0  \textwidth, height= 0.35 \textheight]{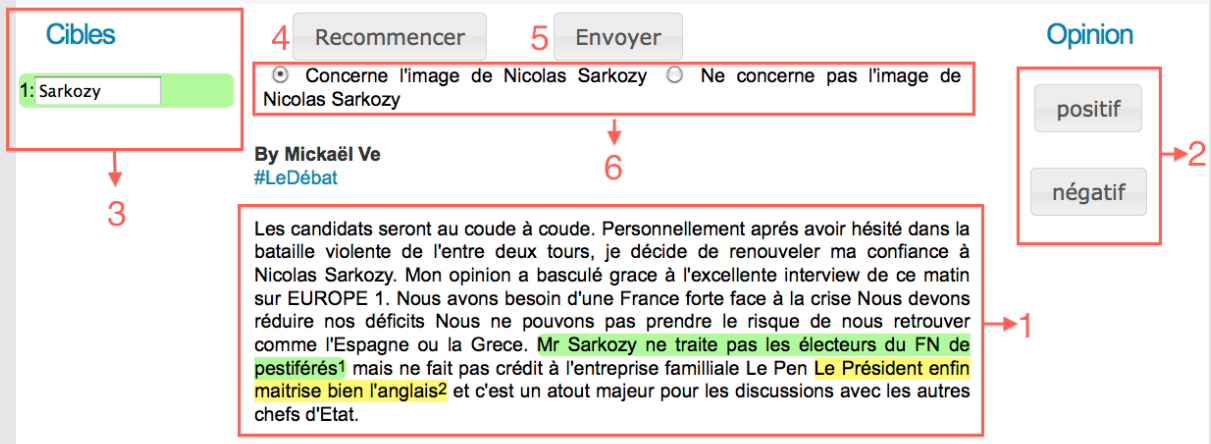}
  \caption{A system for the annotation of tweets polarity in French.}
  \label{figtweet}
\end{figure}

System demo can be tested at \url{http://dev.termwatch.es/~molina/sentaatool/info/systeme_description.html}. Figure \ref{figtweet} shows the interface used during the annotation of tweets and its main components:

1. \textbf{Tweet area} allows selection but not modification.

2. \textbf{Polarity buttons} assign the polarity of a selected passage and make appear a target text bar when pressed.

3. \textbf{Targets section} contains one editable target text bar for each selected passage showing the color depending on the polarity.

4. \textbf{Restart button} restores the interface to initial conditions. 

5. \textbf{Send button} sends the annotations to the database and displays the next tweet to analyze.

6. \textbf{Confidence radio buttons} allow annotators to indicate if the tweet is out of context. Useful if the corpus was extracted automatically.

\subsection{Annotation design}
Designing the set of appropriate aspects is a key element of the whole annotation process. This step has been done under the supervision of experts in political sciences. The following 9 aspects have finally been selected to describe French politicians: attribute \footnote{Poll results and comments}, assessment, skills, ethic, injunction \footnote{Call for voting}, communication, person, political line, project, adding the entity itself and the case of no aspect belonging to this list. The aspects are moreover decomposed into sub-aspects such as polls and support in case of attribute, which signifies the entity's features expressed in pools and supports. At all 23 sub-aspects have been created for this fine-grained description and reporting. The polarity levels vary from very positive (positive) to (negative) very negative opinions, with a neutral opinion (used for facts reports). We also considered an ambiguous opinion for undecidable cases. 

\subsection{First annotated dataset, descriptive Statistics}
Here we provide some statistics about the first dataset (more detailed statistics are available in,~\cite{Velcin2014}). This dataset\footnote{The raw dataset is available there: \url{http://mediamining.univ-lyon2.fr/velcin/imagiweb/dataset.html}} consists of 11527 manual annotations expressing the opinion describing two French politicians over time, 5286 annotations for Fran\c{c}ois Hollande (\textbf{FH}) and 6241 annotations for Nicolas Sarkozy (\textbf{NS}).

Data has been annotated by 20 academics from various fields, Table~\ref{table:anno} provides some additional details. It is interesting to notice that NLP researchers and people from industry focus on terms or N-grams with shorter annotations (in terms of selected passages) probably following respectively algorithm schemes and keywords extraction for dashboards. While at the same time, engineers and politics researchers tend to select larger parts of text.
\begin{table}[!ht]
	\centering
    \small
    \tabcolsep = 2\tabcolsep 
    \begin{tabular}{|c|c|c|c|}
      \hline
      		Domain 							&  Annotators 	& Annotations 	& Average passage length\\
      \hline
      		Computer Science (Engineer)		& 3 			& 2649 			& 90   \\
      		Computer Science (Data Mining)	& 6 			& 1174 			& 82.2 \\
      		Computer Science (IR)			& 3 			& 273 			& 86.5 \\            
      		Computer Science (NLP)			& 2 			& 1747 			& 75.3 \\
     		Energy 							& 2 			& 1070 			& 68.1 \\
     		Politics 						& 4 			& 3407 			& 89.9 \\
      \hline            
            Total 							& 22 			& 11527 		& 82.5 \\
      \hline
    \end{tabular}
    \caption{Number of annotators and annotations from each domain.}
    \label{table:anno}     
\end{table}
To handle the subjectivity of annotators, we allowed a tweet to be annotated at most three times by different annotators. It also happens that the same content (in case of retweet) has been annotated several times by the same annotator which allows us to evaluate the annotator's consistency (details are given below). 7.283 unique tweets (6.369 unique contents) are annotated, of which 48\% are annotated only once, 46\% twice, 6\% three times or more. But, is this enough ? How much are these examples really informative ?

\subsubsection{Opinions} 
For a reasonable analysis, as observed in the literature for comparable annotation tasks,~\cite{carrillo2014orma,villena2013tass}, we consider only three polarity levels by grouping positive/very positive and negative/very negative, and by ignoring ambiguous opinions. On the whole dataset opinions are biased to the negative with a slight difference between the two entities; for example, 47\% of the opinions about \textbf{NS} are negative for 20\% positive and around 53\% are negative for \textbf{FH} while 14\% are positive. The neutral class distribution is equivalent for both candidates with 32\%. However, in the period just before the election (mid-May 2012), the negativity about FH decreases to 41\% while that of \textbf{NS} increases to 52\%. After the election (June to December 2012), the negativity about \textbf{FH} increases dramatically to 72\% as the positivity collapses to 5\%. Per month distributions are summarized in Table~\ref{table:polaritydistrib}. This justifies the necessity of temporal analysis related to the image, with well-split time periods.
\begin{table}[!ht]
	\centering
    \tabcolsep = 2\tabcolsep 
    \begin{tabular}{|c||c|c|c||c|c|c|c|}
      \hline
      		\multicolumn{1}{|c|}{} & \multicolumn{3}{|c|}{ Hollande} &\multicolumn{3}{|c|}{Sarkozy}\\
      \hline
      		Date &  Positive & Neutral & Negative & Positive & Neutral & Negative\\
      \hline
      		March 		& .27 	& .28 	& .45 & .18 & .36 	& .46\\
     		April 		& .25 	& .36 	& .39 & .20 & .34 	& .46\\
     		May 		& .25 	& .34 	& .41 & .19 & .29 	& .52\\
      		June 		& .10 	& .31 	& .59 & .40 & .50 	& .20 \\
      		July 		& .13 	& .35 	& .52 & .24 & .25 	& .50 \\
      		August 		& .8 	& .31 	& .61 & .23 & .30 	& .46\\
      		September 	& .8 	& .32 	& .60 & .26 & .33 	& .42\\
      		October 	& .10 	& .36 	& .54 & .21 & .33 	& .46\\
      		November 	& .7 	& .34 	& .59 & .17 & .32 	& .52\\
      		December 	& .5 	& .23 	& .72 & .20 & .31 	& .49\\
      \hline
    \end{tabular}
    \caption{Polarity distribution across the time on the annotated set from March to December 2012.}    
    \label{table:polaritydistrib}     
\end{table} 

\subsubsection{Aspects} 
The 9 aspects are globally well distributed. As a global class, the entity aspect dominates with 23\%, followed by political line and ethic with 13\ and 11\% respectively. The evolution of the frequency of each aspect according to time is interesting. Some aspects are much more dependent on time such as injunction and communication obtaining very high frequencies just before the election and disappearing after. Both candidates obtained positive opinions for the injunction because this aspect is dedicated to the clear encouragement or warning (rare) about voting for an entity. On the contrary, for the communication FH obtained a better score compared with his competitor. 

\subsubsection{Annotator bias and disagreements} 
The manual annotations may reflect the subjectivity of each annotator because of the granularity of the labels. Despite the task's difficulty, the annotator's low-confidence indicator was only used for 10\% of the annotations and related to the "ambiguous" polarity level. As it was not properly used, we reconsidered the quality of mainly non-expert annotations on different aspects. While for a machine a word sequence will match a unique model or a weighted number of models, the language acquisition skills of humans result to a multidimensional experience. Then, annotating is dependent from annotator's language acquisition skill. Analyzing the annotation disagreement among annotators for each tweet would provide us with a better understanding of the opinion properties. Considering this, we try to analyze the problem at the content level by taking a closer look at the annotations from the text level. We can now observe more severe disagreement for a unique content since annotators may have different backgrounds and points of view on the same document

We assume here that we do not need to explore the idea of recalibrated annotator judgment to more closely match expert behavior or to exclude some annotators from the process. If polarity disagreement is less than 20\%, disagreements on aspects (including sub-aspects) exceed 60\% with a basic analysis. Things get worse when considering the cascade, disagreements increase dramatically on the polarity-aspect. This is explained on one hand by natural language variability, the background knowledge of the annotator that may make him interpret a hidden meaning of the message while others did not notice the irony. On the other hand it comes from the concept variability. It can be illustrated with the case 'Sarkozy-Kadhafi' which has been correctly tagged as ethic by the two annotators, but the chosen sub-aspect differs (ethic: honesty vs. ethic: case). A typical example for polarity, despite the guidelines, can be a tweet that describes the result of a public poll. If the poll is in favor of a candidate, some annotators give a positive (resp. negative) polarity while others give a neutral polarity since they consider this information as a fact. 

Things become interesting when looking how annotators labeled a repeated content. For each content annotated more than five times we can observe that there is on average one annotation different from the others, annotator's consistency is estimated to be around 80\%. It illustrates the fact that different aspects can be selected depending on the individual point of view but also offers us the possibility to see the trends of an annotator. As the annotation stage lasted over several weeks it will be subject to variation.

\section{Intelligent annotation framework} 
\label{sec:data}
We described within this study how we exploit text mining techniques to analyze a real-world data sample from Twitter. As mentioned before, one difficulty is to have enough data and information to build models for employing machine-learning approaches. Despite the recent advances and good practical results, improvements remain to be achieved. "How much is enough ?" is still an open question. Our main objective is to bootstrap machine learning techniques using limited annotated data to detect how a given entity is perceived. Then, to follow-up Active Learning minds and enhance data informativeness on the time, we also experiment with some approaches to apply recommendation system's adaptation to re-build models on the fly.

An important fact is that this public perception is not static and may change in time which implies to adapt models. However, experts in political science need a huge amount of tweets to release a deep, complete and reliable analysis over time. Therefore, getting involved on such an annotation campaign is not possible in financial terms. In our case it has been decided that both NLP and political researchers will work jointly in a pseudo-active learning process. To achieve this objective, we set a semi-automated step which aims at evaluating the quality of text mining technique submissions. These automatic suggestions are then compared to real-world results, that is to say expert committee decisions to validate our algorithms.
\begin{algorithm}[!ht]
 \KwData{Large amount of unlabeled tweet}
 \KwResult{Large amount of labeled tweet}
 Small amount of tweets are manually labeled\;
 \While{Not enough labeled data or insufficient classifiers' performance}{
  Build models with the labeled data\;
  Classify a subset of unlabeled data\;
  Select and send a sample of automatic classification outputs for manual confirmation\;
  \eIf{Automatic classification is sufficient}{
   Annotation for the whole dataset\;
   }{
   Go back to the beginning with more labeled data for learning\;
  }
 }
 \caption{Annotation process.}
\end{algorithm}
These choices have been submitted to validation through a couple of experiments (see Section~\ref{subsec:result}).

\subsection{Data Diversity} 
Our objective is then to train machine learning techniques using human behaviors in order to propagate their knowledge and automatically label forthcoming data. Something important before handle a large unannounced data set is to be sure about the training set reliability. As reported by the literature,~\cite{artstein2008inter} and as we have just seen, human annotations of language features and concept are prone to human errors. These errors need to be considered in the model learning process since it is well known that the quality of manual annotation is critical when it comes to train automatic methods. We assume that the objective is not to build the most reliable dataset in the meaning of a particular aspect but to build a consistent dataset regarding to the upcoming analysis we want. For the training step, instead of misleading the automatic algorithms, we can consider that this situation will reflect the diversity of interpretation. We could consider that a message conveys two messages (e.g. two topics with each a different opinion) by the multi-label,~\cite{tsoumakas2006multi}) but we made the questionable assumption that one tweet = one opinion and one topic. Because when it comes to the evaluation set, it is critical to agree about only one reference. When working with learning and data mining on text contents we have to keep a high variability in the data distribution (in terms of contents and labels) to prevent falling into a biased distribution that will lead us in over-training and then to overvalue our systems. It is also difficult to distinguish the real informative examples (from the non-informative ones), and the fact that it will only be possible to annotate content similar to labeled ones is an important drawback. Regarding the cost of a such annotation stage we need to maximize the effectiveness of each annotation by having certified label on the largest vocabulary as possible. This step can be seen as text processing since each content is cleaned in order to detect duplicate messages and ignore them for the further annotation steps. For a more focused work on aspects (e.g. statistics about voting) we keep a track of all duplicates in order to propagate the annotation.

\subsection{Harmonization}
\subsubsection{Annotators-based decisions}
It is still possible to estimate the task difficulty with inter-annotation agreement measures such as Kappa,~\cite{kohen1960coefficient,cohn1994improving,Koehn2003,sabou2014corpus} but once disagreements have been identified what can be done ? In our case, each tweet has been annotated from one to three times and as we before noted severe disagreements at text level we have chosen a majority-based rule system. For each annotated content with divergent annotations, we selected, whenever possible, the human annotation that has a relative majority according to:
\begin{center}
  \begin{equation}
     \textrm{Label frequency} > 1/\textrm{Number of labels}
  \end{equation}
\end{center}

\subsubsection{Profiles-based decisions}
\begin{figure}[!ht]
  \centering
  \includegraphics[scale=0.5]{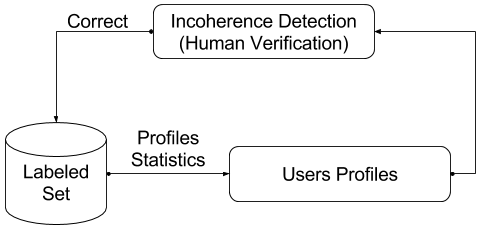}
  \caption{Annotation errors corrections with users profiles.}
  \label{fig:profile_correction_base}
\end{figure}

For a given tweet, none of the labels has the majority therefore we have chosen to work at the user level (as shown in figure~\ref{fig:profile_correction_base}). 

An important aspect in social networks is the possibility for users to answer each other thus building their own network. We can consider as an extra feature that a user belongs to a group or has the same opinion (or aspect) as the person to whom he or she responds or re-tweets. Moreover, considering the political dimensions of the data set, we assume that in a short time period gossipers expressing their revulsion about one candidate cannot find something positive in only one message. We then need to pay attention to these annotations. For instance, we can consider users having more than 100 negative messages related to a given entity. We can hardly imagine the next tweet to be positive and even if it has been annotated as such, it may be withdrawn or submitted to a new validation. This process can be seen as smoothing the user point of view, even if we know that this is an assumption is not always verified. Some NLP analysis has also been considered with a few nicknames such as \@nainportekoi (\@dwarf + anything with bashing) or \@hollandouillette (contraction between Hollande and sausage which also means stupid).

Although this method might be the first step towards specific processing for polarity, we are not able to apply it on the aspect classification task, since tweets' authors are not only talking about one specific aspect. A similar method can then be considered for hashtags since it has been proven that hashtags often carry specific topic information ,~\cite{BrunRoux14}.

\subsubsection{Contents-based decisions}
We also investigated contents-based correction making use of the statistical information. In particular we first investigated sentiments carried by hashtags, e.g. \#LesSocialos is always associated with negative opinion about FH, tweets containing this hashtag annotated as positive should draw attention. Hashtags are used to label groups and topics on Twitter; they can be categorized into three types:
\begin{itemize}
  \item Topic hashtags, used to annotate coarse topics, e.g. \#LeDebat (\#TheDebate) \#Karachi (case);
  \item Sentiment hashtags, e.g. \#Idiot (\#Idiot), \#Deception (\#Disappointment), \#LesSocialos (\#socialists - with bashing) and various stylish forms of umpitoyable, umpitres, umpopcorn (bashing UMP party);
  \item Sentiment-topic hashtags, which captures both sentiment and target topic, e.g. \#ViveHollande (\#LongLiveHollande), \#SarkoOnTaime (\#SarkoWeLoveYou) \#Nabot (\#dwarf with bashing).
\end{itemize}

Then, in addition to hashtag, we considered statistical NLP,~\cite{sparck1972statistical,salton1988term} with N-grams to compose the tweet discriminant bag-of-words (BOW) representation using normalized, inverse term frequencies (tf-idf),~\cite{robertson2004understanding} and Gini criterion,~\cite{cossu2015nlp,torres2013bechet}. We consider that a tweet requires additional attention when the most discriminant terms it contains are not corresponding to its label. For instance we used the statistical information to correct annotations considering terms such as: "au-secours sarko revient" ( Help Sarko is coming back), "sarkocasuffit" (Sarko that is enough) directly and negatively related to NS. Rather than considering french domain-specific lexicons such as those mentioned by,~\cite{smeaton1999using,pla2014political} for English and Spanish, this approach is more flexible and requires less resources. 

\section{Setting-up machine learning framework : Issues and Challenges} 
\label{sec:classif}

\subsection{Machine Learning Committee-based correction}
Unlike,~\cite{dagan1995committee}, we consider a different committee-based validation composed by several classifiers which are described above under the very light supervision. Domain non-specialist check different random samples of system outputs to validate the process. Some studies worked well in the first direction, such as,~\cite{liere1997active} where the authors obtained 2- to 30- fold reductions of the amount of human annotation needs for text categorization. 

\begin{figure}[!ht]
  \centering    
  \includegraphics[scale=0.5]{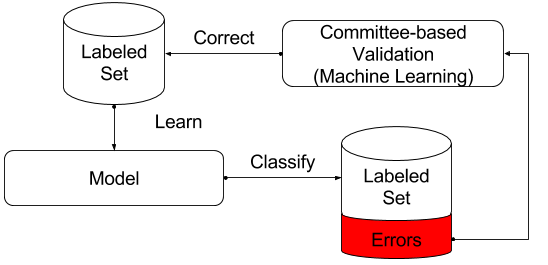}
  \caption{Annotation errors detection with machine learning.} 
  \label{fig:machine_correction}
\end{figure}

After the rules-based corrections, for all remaining cases we now resort to several classifiers used to "self annotate" the training corpus. A wide number of methods have already been explored to correct the bias of annotators. Having multiple annotators is a case that we allow, however an important fact here is that we do not consider annotations as gold standard reference and we can question them especially if none of them has a true label of the systems agreed on. We assume that classifier outputs can be considered as several additional referees for a committee-based validation at the same level as human annotators (as described in figure~\ref{fig:machine_correction}) in different way such as leave one out process.
In the self annotating corpus, we observed that for the original set classifiers are not able to find the correct label for a part of the set. For instance with the cosine distance Accuracy and $\mu$ F-Score to be respectively $.84$ and $.87$ for FH, $.84$ and $.83$ for NS. From these classification errors we distinguished several cases : 
\begin{enumerate}
  \item All system agreed on a label different from the human annotations;
  \item A majority agreed on a label different from the human annotations;
  \item No agreement;
\end{enumerate}
Based on the majority rule expressed above, we now consider for the two first cases (around 60\%) the prediction of the classifiers as the new ``reference'' annotation for the tweet. In the last case tweets are submitted again for human verification. It is interesting to notice that except for some ironic tweets, after the correction classifiers are now able to find the correct label for a very high majority of tweets obtaining more than $.98$ in each measure.

\subsubsection{Classifiers}
For the purpose of this experiment and following the background literature,~\cite{cossu2015nlp}, we investigated statistical NLP,~\cite{sparck1972statistical,salton1988term}. N-grams also compose the tweet discriminant bag-of-words (BOW) representation using normalized (tf) inverse term frequencies (tf-idf) and Gini criterion,~\cite{cossu2015nlp,torres2013bechet}. The statistical BOW approach is used to compute the similarity of a given tweet to each class BOW and rank tweets according to Jaccard index, cosine distance and the score provided by several classifiers (Poisson-based classifier, Hidden Markov Model) ~\cite{Cossu2013LIARepLab2}. We also proposed a kNN-based classification method that uses the same discriminant factor as the one used in the BOW representation. We match each $d$ document from the test collections to the K-most similar $d$ documents in the training set using Jaccard index and cosine distance to measure document similarity. The K most similar tweets vote for their class according to their similarity with the tested tweet. 

Rather than selecting the best hypothesis we considered all output scores provided by classifiers for each class. Then all scores have been normalized, between 0 and 1, so that they can be merged considering a linear combination, weighted linear combination and multi-criterion optimization methods,~\cite{lamontagne2006combining,batista2012multi}. The combination procedure follows two rules:
\begin{enumerate}
  \item maximize the confidence of automatic annotation by using combined classifier scores, 
  \item follow the label distribution observed in the training set.
\end{enumerate}
We consider a specific combination for each entity and sub-task (polarity or aspect).

\subsubsection{Metrics}
The absolute values from confusion matrix are used to calculate usual text mining metrics as Accuracy. Which although it is easy to interpret, it is nevertheless easy to be cheated under unbalanced test sets. For instance, a non-informative method returning all tweets in the same class (all ``\textit{NEGATIVE}'' in our case), may have high accuracy. We also compute an average F-Score, based on Precision and Recall for each class, typical in categorization tasks which is calculated as follows:
\begin{equation}
	\textrm{F\_Score} = \dfrac{\sum\limits_{c} \dfrac{2 \times (\textrm{Precision}_c \times \textrm{Recall}_c)}{\textrm{Precision}_c + \textrm{Recall}_c}}{\text{Number of classes}}
\end{equation}

\subsubsection{Datasets}
We divided the corpus into two parts, chronologically sorted: training (Tr) and development (D). D was built with the 3 last months (approx. 800 unique contents associated with each entity). 

This initial subset has been extended to more unlabeled tweets extracted from Jan. 2012 to Dec. 2014:
\begin{itemize}
\item A first set concerning FH containing 240k tweets (around 6700 tweets per month)
\item A second set concerning NS containing 81k tweets (around 2500 tweets per month)
\end{itemize}
This new data is used for the validation process and the experts need them for drawing conclusions at large scale by using the prototype. Around 3000 tweets have randomly been selected each month over 21 months from January 2012 to December 2013 which led to 51020 unique contents for FH (and 16050 for NS) to provide background context for systems. All tweets from 2014 will form our validation set which will be reviewed by experts (see below). 

\subsubsection{Integrating users information}
\begin{figure}[!ht]
  \centering
  \includegraphics[scale=0.4]{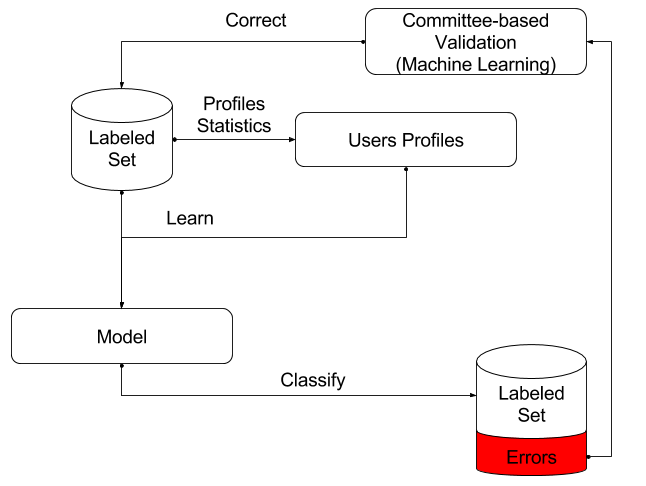}
  \caption{Combining Machine Learning and users profiles in the annotation correction process.}
  \label{fig:profile_correction}
\end{figure}

For the users concerned by profiles-based annotation corrections, we considered a smoothing in the machine learning approaches (as summarized in figure~\ref{fig:profile_correction}). We first added a class tag in the bag-of-words of the future tested tweets (which represents the main polarity they were associated with, by the classifiers in the BOW of their tweets). Nevertheless this tag implies that the user will not change his mind. To prevent this bias and also accept that people can change their mind without breaking the BOW robustness, we then added the user identifier with its associated classes' probabilities,~\cite{li2011incorporating}. This way, by looking at the past of this user, we penalize the contribution the non-majority class without closing doors to a further change in user's mind. Since, we are in an Active process, as time goes on it will automatically return on the premise that one user has only one opinion.

\subsection{Wrap-up}
Table~\ref{table:corrml} summarizes the corrections made. Although NS only possess 17\% additional raw annotations regarding FH, it concentrates much more corrections regarding the opinions while conversely the trend is reversed with respect to the aspects. We can mainly explain this with the label distribution since the positive classes are not really existing with FH, it lower the task's complexity. ML and content-based approaches did not help much to improve the annotation-correction process for the opinion detection issue while profile statistics appeared to play a key role. In addition, it is interesting to notice that for NS even after a committee statement it was still impossible to agree on a label for some messages which were finally rejected. Finally in many cases regarding aspects, neither rules, ML approach nor the committee were able to agree and an additional referee was asked to provide a supplementary annotation. 
\begin{table}[!ht]
	\centering
    \tabcolsep = 2\tabcolsep 
    \begin{tabular}{|c|c|c|}
      \hline            
      		Correction Type 			&  \multicolumn{2}{|c|}{\# of correction} \\    
      \hline
      		Polarity					&  Hollande & Sarkozy \\
      \hline
      		Contents-based				& 30 		& 15 \\
      		Rules-based with annotators	& 15 		& 71 \\
      		Rules-based with nickname	& 141		& 446 \\
      		Rules-based with ML			& 24 		& 25 \\
      		Rules-based with Hashtags	& 13 		& 38 \\            
      		Committee-based				& 101		& 411 \\
      		Reject						& 1			& 9 \\                        
      \hline
      		Total						& 324		&  1015 \\                  
      \hline
      \hline
      		Aspects						&  Hollande & Sarkozy \\
      \hline
      		Contents-based				& 0			& 28 \\
      		Rules-based with annotators	& 885		& 372 \\
      		Rules-based with nickname	& 0			& 3 \\
      		Rules-based with ML			& 103 		& 217 \\
      		Rules-based with Hashtags	& 0 		& 6 \\            
      		Committee-based				& 94		& 25 \\      
      		Reject						& 0			& 61 \\      
      		New annotation				& 349		& 297 \\            
      \hline
      		Total						& 1401		& 1009 \\            
      \hline
    \end{tabular}
    \caption{Numbers of corrections for each candidate and each task.}
    \label{table:corrml}     
\end{table}

Table~\ref{table:annorecap} summarizes the correction with regard to the annotators groups. It is interesting to note several points. We can observe a major difference between NS and FH, in the first case (NS), there are more opinion corrections (since it is a 3 class problem while FH holds two polarity levels having only poll and injunction as positive examples). Whereas in the second case FH holds much more mistakes on aspects mainly concentrated between assessment, political line and project but also between skills and communication. For NS aspects mistakes appear to be limited between ethic and person. Annotations on tweets concerning NS presents more stability between aspect and opinion with similar error rates.

Concerning groups of annotators, there are several tiers, for aspects with FH, engineers are leading while politics and IR researchers missed something. Conversely, with regards to opinion, IR researchers made less mistakes doing even better than politics. The situation is quite different with NS, because error rates for opinions are quite similar between groups with a lead for engineers. However for aspects, IR researchers group still obtain the lower results and politics fall short with the lead.

\begin{table}[!ht]
	\centering
    \small
    \tabcolsep = 2\tabcolsep 
    \begin{tabular}{|c|c|c|c|c|c|}
      \hline    
		    \multicolumn{6}{|c|}{\# Hollande - FH} \\    
      \hline
      		Domain 				&  Annotations  		& Aspect corr. 		& Polarity corr. 				& Corr. Rate  					& Corr. Rate \\
      \hline
      		C.S. (Engineer)		& 1298 					& 294 					& 71 							& .227 						&  .055 \\
      		C.S. (Data Mining)	& 982 					& 246 					& 80 							& .251 						&  .081 \\
      		C.S. (IR)			& 89 					& 33 					& 3 							& .371 						&  .034 \\
      		C.S. (NLP)			& 741 					& 243 					& 38 							& .328 						&  .051 \\
   			Energy 				& 536 					& 158 					& 29 							& .295 						&  .054 \\
   			Politics 			& 1640 					& 446 					& 104 							& .328 						&  .051 \\
      \hline
      		Total				& 5286 					& 1420 					& 325 							& - 						&  - \\
      \hline      
      \hline    
		    \multicolumn{6}{|c|}{\# Sarkozy - NS} \\          
      \hline
      		Domain 				&  Annotations  		& Aspect corr. 		& Polarity corr. 					& Corr. Rate  				& Corr. Rate \\
      \hline
      		C.S. (Engineer)		& 1351 					& 207 					& 188 							& .153 						&  .139 \\
      		C.S. (Data Mining)	& 1632 					& 266 					& 312 							& .163 						&  .191 \\
      		C.S. (IR)			& 182 					& 43 					& 30 							& .236 						&  .165 \\
      		C.S. (NLP)			& 772 					& 135 					& 106 							& .175 						&  .137 \\
   			Energy 				& 534 					& 100 					& 99 							& .187 						&  .185 \\
   			Politics 			& 1767 					& 250 					& 272 							& .141 						&  .154 \\
      \hline
      		Total				& 6238 					& 1001 					& 1007 							& - 						&  - \\            
      \hline      
    \end{tabular}
    \caption{Overview of annotations and corrections within each groups of annotators. C.S. stands for Computer Science.}
    \label{table:annorecap}     
\end{table}

After all changes introduced by the process described above, the polarity distribution from the original set can be altered. In the period after the election (\textit{June to December 2012}), the negativity about \textbf{FH} increases dramatically to 79\% near the end of the year while there was only 5\% left in positivity. We then study the impact of the harmonization process described above on the results of classifiers on the last tweets of the dataset considered here as test set (as if we were simulating incoming data or temporal expansion). In other words, we consider improvement in the output of polarity assignment to evaluate the gain offered by the harmonization process. cosine performances then increased for both FH and NS from respectively F-Score and Accuracy $.37$, $.60$ to $.44$, $.69$ and $.40$, $.46$ to $.43$, $.51$ the relatively small size of the test set did not permit us to compute significance test. In a next experiment we have annotated the large set of unlabeled tweets and considered these new annotated data as new training material and retried polarity assignment on our small test-set. 
Regardless of its size, this training set may not be completely reliable, performances for FH respectively reached F-Score and Accuracy $.46$, $.66$. Moreover, observed improvements for positive and neutral tweets prove that our propagation do contain relevant information that improves the polarity classification and that was missing on the original set.

\subsection{Expansion, temporal propagation}
\begin{figure}[!ht]
  \centering
  \includegraphics[scale=0.5]{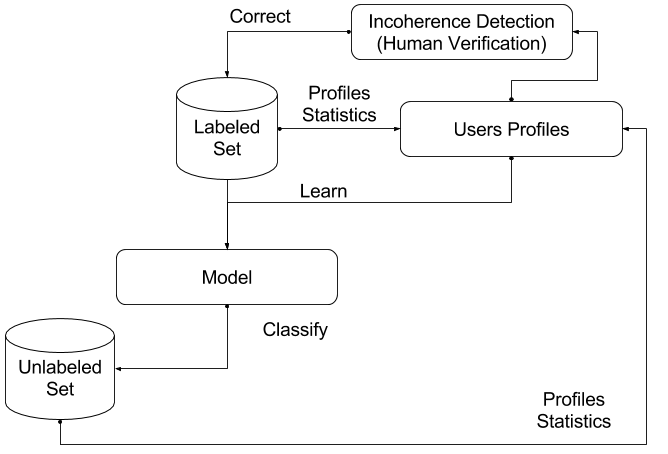}
  \caption{Temporal expansion to improve users profiles.} 
  \label{fig:utp}
\end{figure}

Now that the training set has been corrected, we can use our classifiers to annotate a large set of unlabeled messages (as summarized in figure~\ref{fig:utp}). 
The unlabeled examples can be used with unsupervised or supervised learning methods to improve the classification performance and the correction of the labeled examples by applying the above rules according to a principle of homogeneity at content and user level.

\begin{figure}[!ht]
  \centering
  \includegraphics[scale=0.5]{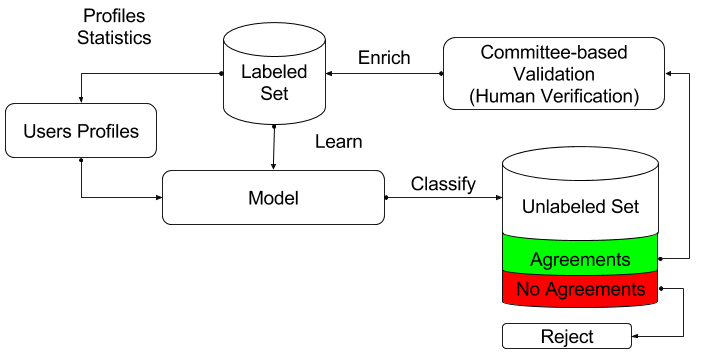}
  \caption{Outliers Detection Process.} 
  \label{fig:outliers}
\end{figure}

Additionally we considered 'outliers' which are examples that differ from the rest of the data. In our case in terms of agreement or content. We first considered \textbf{excluded-outliers} as tweets that neither systems or annotators agreed on the same label. These tweets will be ignored because of understanding shortages. We also excluded unique contents with no common words with other contents and with the labeled set. A second interpretation of \textbf{reliable-outliers} is to respectively consider tweets for which every system agreed on the same label  by adding them in the labeled set before iterating,~\cite{spina2015active}. These 'reliable-outliers' were verified by human annotator which agreed on automatically chosen label. After this step, as we consider them reliable enough to be used as models, these tweets were no longer candidates for a next manual annotation step (as shown in figure~\ref{fig:outliers}). 

\section{Evaluation}
\label{sec:results}

\subsection{Evaluation data}
\begin{figure}[!ht]
  \centering
  \includegraphics[scale=0.5]{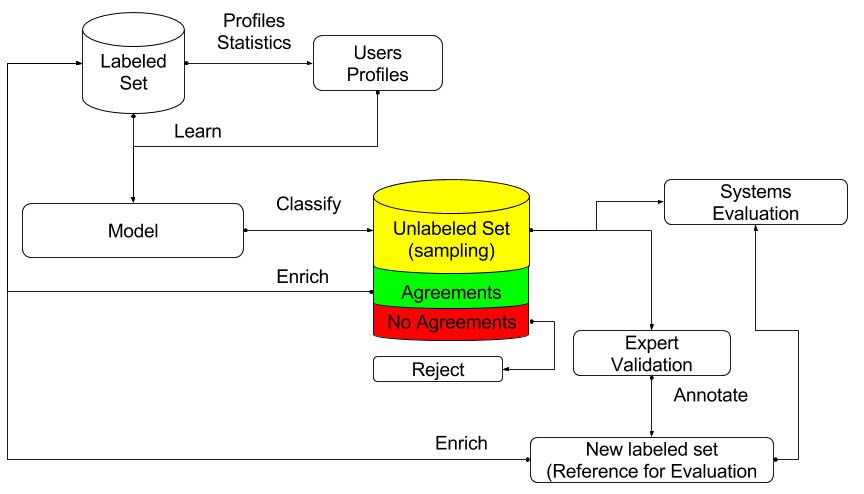}
  \caption{Summary of the evaluation process.} 
  \label{fig:evaluation}
\end{figure}

We consider as test data set a selection of 5200 tweets in 2013 (430 each month) for NS and 3600 tweets (March and April 2013) for FH. These selected tweets were automatically annotated with the workflow presented below and were also manually reviewed by an expert in political science following the annotation guidelines (as summarized in figure~\ref{fig:evaluation}). Note that, for the entity NS, we divided the set in two parts: a first one where the automatic label was completely hidden to the annotator (similarly to raw tweets annotation), and a second one where the automatic label was shown to the annotator (validation/correction stage if it was wrong). The test set of entity FH was validated following this second scheme. Below, we compare the expert annotation with the hypotheses automatically produced. The goal of this setup is twofold, first we intend to evaluate the performance of machine learning approach in an operating scenario. Secondly, we want to estimate how much an annotator can be influenced (or not) by automatic suggestion during the validation step.

\subsection{Results}
\label{subsec:result}
As preliminary experiments, we first report in Table~\ref{tab:resopi} the system performances for the classification tasks (polarity and aspect respectively) on the two studied entities on our test sets. To keep things simple we only report the performances of a cosine-based approach and the combination of all machine learning techniques used during the annotation process. Although there is a significant improvement in the evaluation of the classification, the most important is that the combination of classifiers also appear to be robust enough to handle the large variety of hypotheses.

Then, regarding the fact that the annotator was able to see the automatic label (or not for half of NS tweets) when he was annotating the tweets, differences are not significant for the polarity classification (Accuracy between $.62$ and $.63$ for combination of classifiers). Although as the task of annotating the tweet according to only one aspect is difficult, so we can consider that the annotator validated the proposed aspect by convenience because it was not so wrong even if there could have been another possible choice. F-S for $Combination$ was situated at $.25$ when the annotator was not able to see the automatic label, and $.33$ in the second case, given the number of aspects the difference is quite significant.

\begin{table}[!ht]
  \centering
  \tabcolsep = 2\tabcolsep
  \begin{tabular}{|lcc||cc|}
    \hline
	    \multicolumn{1}{|l|}{Opinion} & \multicolumn{2}{|l||}{FH} & \multicolumn{2}{|l|}{NS}    \\
    \hline
    	Sys & \textbf{F-S} & Acc & \textbf{F-S} & Acc \\
    \hline
	    Cosine & .535 & .754 & .504 & .617 \\
    	Combination & .535 & .757 & .520 & .620 \\
    \hline
    \hline    
    	\multicolumn{1}{|l|}{Aspects} & \multicolumn{2}{|l||}{FH} & \multicolumn{2}{|l|}{NS}    \\
    \hline
    	Sys & \textbf{F-S} & Acc & \textbf{F-S} & Acc \\
    \hline
    	Cosine & .367 & .468 & .280 & .463 \\
    	Combination & .369 & .473 & .269 & .451 \\    
    \hline        
  \end{tabular}
  \caption{Systems performances in terms of F-Score with entity-specific models for opinion classification and global models for aspects classification.}  
  \label{tab:resopi}
\end{table}

In additional experiments, we tried to switch and combine entities models. That is to say predict NS polarity using NS, or FH or FH+NS training set. The aim of these experiments is to test how well the method can perform without proper training material or with opposite sentiment. For the polarity classification, the results were obviously lower with combined and switched models than with the entity specific models. Trying to classify FH tweets with FH models leads to a F-S value of $.52$ and around $.66$ Accuracy while FH+NS models stay a bit lower with respectively $.48$ and $.64$, considering only NS models. Cf. performances collapsing at $.41$ for both metrics. Indeed, in terms, for example, of political \texttt{balance sheet} and \texttt{project}, what can be seen as a positive statement about one candidate may be rather negative for the opposite side whereas it is expressed with the same words. Conversely, we have considered that aspects do not depend on a specific entity but have a consistent cross-entity behavior. Consequently, we considered both entities altogether to address the aspect-oriented classification issue. Combined models appeared to be a semantic enrichment and show a slight improvement in classification performance for both entities. This led us to then consider and report only combined models performances.
    
\section{Conclusion and perspectives}
Depending on the domain it is applied on Sentiment Detection. This task is even more difficult when it comes to combine it with the specific aspects. In this paper, we presented an approach to annotate a French political opinion dataset from annotation design to machine learning experiments. 

First, we have shown that we can improve our dataset and obtain good classification performances even though statistical methods are without linguistic and domain specific processing. That makes our approach easily applicable to other languages and dataset. Instead of addressing a more complex modeling, experiments reported in this paper have shown that by considering additional Twitter Features combined to light knowledge, this can provide a robust support to improve both annotation quality and classification performance. 

We employed methods known to remain simple but also reported to obtain results as good as the ones proposed so far with the state-of-the-art approaches on comparable issues,~\cite{cossu2015nlp}. We demonstrated our approach efficiency by comparing automatic aspect-oriented opinion annotation of tweets to label that have been proposed by experts in political science. 

As the need for in-domain annotated data still persists we hope that the methods and tools presented here will help researchers in their quest of bigger and better dataset. Solving this problem could help prevent from annotator bias and errors and minimize human oversight, by implementing more sophisticated computer-based annotation work-flows, coupled with in-built control mechanisms and low supervision. Such infrastructure needs to be reusable. Further on, we would like to extend our approach on simultaneously predicting the polarity and the aspect it is associated with.

\section*{Acknowledgments}
This work was funded by French ANR project ImagiWeb (under ref. ANR-2012-CORD-002-01) and Ministère de Sciences du Mexique, Conacyt (founding 211963). The authors would like to thank Dr Eric Sanjuan, Pr Marc El-Beze and the whole Imagiweb team especially Dr Caroline Brun.

\bibliographystyle{jimis-en}
\bibliography{jimis-refs}

\appendix\footnotesize

\end{document}